\documentclass[prb,twocolumn,nopacs,nofootinbib]{revtex4-2}
\usepackage{amsmath}
\usepackage[dvipsnames]{xcolor}
\usepackage{nicefrac}
\usepackage{graphicx}
\usepackage{microtype}
\usepackage{physics}
\usepackage{mathrsfs}
\usepackage{amssymb}
\usepackage{bbold}
\usepackage{ dsfont }
\usepackage{pifont}
\usepackage{multirow}
\usepackage[bookmarks=true,colorlinks,linkcolor=blue,urlcolor=blue,citecolor=blue]{hyperref}
\usepackage[english]{babel}
\usepackage[applemac]{inputenc}
\usepackage[T1]{fontenc}

\newcommand{\beq}{\begin{equation}}
\newcommand{\eeq}{\end{equation}}
\newcommand{\beqa}{\begin{eqnarray}}
\newcommand{\eeqa}{\end{eqnarray}}

\usepackage{soul}
\usepackage[normalem]{ulem}

\graphicspath{{../}}

\begin{document}

\title{Multiple non-hermitian phase transitions on quantum torus surface}
\author{Jos\'e A. S. Louren\c{c}o$^{1}$ and Ygor Par\'a$^{2}$ and J. Furtado$^{3,4}$}

\affiliation{
$^{1}$Departamento de F\'isica, Universidade Regional do Cariri - 57072-270, Juazeiro do Norte, Cear\'a, Brazil\\
$^{2}$Departamento de F\'isica Te\'orica e Experimental, Universidade Federal do Rio Grande do Norte, 59072-970 Natal-RN, Brazil\\
$^{3}$Universidade Federal do Cariri (UFCA), Centro de Ci\^{e}ncias e Tecnologia, Juazeiro do Norte, CE, 63048-080, Brazil\\
$^{4}$Department of Physics, Faculty of Science, Gazi University, 06500 Ankara, Turkey
}

\date{\today}

\begin{abstract}
In this paper we investigate the arising of non-hermitian phase transitions on quantum torus surfaces. We consider a single fermion whose dynamics is governed by the Dirac equation confined to move on a quantum torus surface. The effects of the geometry are take into account by using the tetrad formalism and the spin connection. The Dirac equation gives rise to two coupled first-order differential equations for each spinor component. The eigenvalues and eigenfunctions for each spinor component are computed numerically and the non-hermitian phase transitions are investigated in terms of the geometric features of the torus and the magnitude of the imaginary component of the mass.   

\end{abstract}

\maketitle

\section{Introduction}\label{sec:intro}

The study of quantum mechanics on curved surfaces reveals captivating phenomena. Quantum particles, even on curved surfaces like tori \cite{GomesSilva:2020fxo} or catenoids \cite{SILVA2020126458}, display wave-like behaviors such as diffraction and interference. The curvature of these surfaces introduces geometric factors that significantly impact quantum particle behavior, leading to intriguing effects like quantized energy levels and geometric phases \cite{dacosta1, dacosta2}. Furthermore, surface curvature can exert curvature-induced forces on particles, altering their trajectories and dynamics \cite{dacosta1, dacosta2}. This exploration of quantum mechanics in curved spaces sheds light on fundamental aspects of both quantum theory and geometry.

Two-dimensional nanostructures like graphene \cite{katsnelson_2012, Geim:2007aeb, CastroNeto:2007fxn} and phosphorene \cite{Phosphorene} are notable in low-energy physics due to their unique properties, heavily influenced by geometry \cite{DACOSTA, COSTAFILHO2021114639, Vanderley}. They serve as analog models for higher-energy physics systems \cite{deSouza:2022ioq, Geovas, IORIO,IORIO20111334, Capozziello, CVETIC20122617, Behnam}. Among these structures, the torus stands out for its significant curvature effects \cite{GomesSilva:2020fxo, Ye_and_Job2022}, with carbon nanotori finding applications in nanoelectronics, biosensors, and quantum computing \cite{biosensors, PhysRevLett.97.016601}.

Studies on toroidal surfaces delve into various aspects. For instance, analyzing electron dynamics with the Schr\"{o}dinger equation yields insights into curvature-induced bound-state eigenvalues and eigenfunctions \cite{Encinosa}, also considering external fields \cite{GomesSilva:2020fxo}. Investigations extend to charged spin $1/2$ particles governed by the Pauli equation on toroidal surfaces \cite{SCHMIDT201988} and analytical solutions for the $(2 + 1)$ Dirac equation on tori, exploring cases with constant or position-dependent Fermi velocity \cite{Ye_and_Job2022}, including scenarios with external fields \cite{Ye_and_Job2022}. Recent work explores qubit encoding using the energy levels of graphene nanotori \cite{furtado2022encoding}. 

Non-Hermitian systems have been highlighted in several areas of physics in recent years. In particular, non-Hermitian systems symmetrical simultaneously by parity $\left(\mathcal{P}\right)$ and time-reversal $\left(\mathcal{T}\right)$ can present the phase transition $\mathcal{PT}$ \cite{Bender}. The $\mathcal{PT}$-transition indicates the spectrum of the system starting from the real eigenvalue and becoming complex or purely imaginary, and can also occur from the complex/imaginary to the real \cite{Bender2007}. The $\mathcal{PT}$ phase transitions are related to critical points (called exceptional points) that limit the $\mathcal{PT}$-symmetric and $\mathcal{PT}$-broken phase transitions \cite{Rotter,Heiss2012,Wei,PhysRevB.98.085126}. Multiple non-Hermitian phase transitions were investigated in the non-Hermitian dynamics of a mesoscopic system of long-range interacting particles \cite{Lourenco22}, Floquet quasicrystals system \cite{Zhou22}, coupled acoustic cavities with asymmetric losses \cite{Ding16}, anisotropic exciton polariton pairs \cite{Chakrabarty23} and non-Hermitian sensing in the control of quasi-parametric amplifications \cite{Wu24}.

Non-Hermitian Hamiltonians appear naturally in curved space as the quantum Beltrami surface \cite{FURTADO2023Electronic}, generalized Ellis-Bronnikov graphene wormhole-like surface \cite{deSouza:2022ioq}, torus \cite{GomesSilva:2020fxo} and catenoid \cite{SILVA2020126458,SILVA2021} which exhibits $\mathcal{PT}$ symmetry in the Hamiltonian. Also, non-Hermiticity can be induced through an imaginary Dirac mass term \citep{Wang,Nori,Ygor2021}. In particular, on the surface of the sphere \cite{Ygor2021}, there is an infinite sequence of exceptional points (EP), which depend on the radius (curvature) that induces non-Hermitian phase transitions. 

In this paper we investigate the arising of non-hermitian phase transitions on quantum torus surfaces. We consider a single fermion whose dynamics is governed by the Dirac equation confined to move on a quantum torus surface. The effects of the geometry are take into account by using the tetrad formalism and the spin connection. The Dirac equation gives rise to two coupled first-order differential equations for each spinor component. The eigenvalues and eigenfunctions for each spinor component are computed numerically and the non-hermitian phase transitions are investigated in terms of the geometric features of the torus and the magnitude of the imaginary component of the mass.  

This paper is organized as follows: In the next section we present the model under consideration and obtain the Dirac equation for the system by taking into account the effects of the geometry. In section III we discuss numerically the non-Hermitian phase transitions for the system and in section IV we highlight our conclusions. 

\section{Model}

\begin{figure}[t]
\includegraphics[width=0.9\columnwidth]{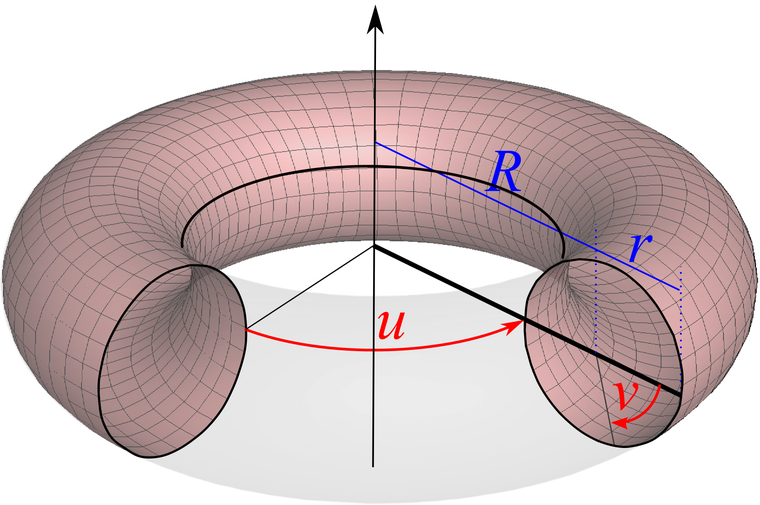}
\caption{Torus surface and coordinate system. \label{fig1}}
\end{figure}

We consider the dynamics of relativistic fermions constrained to the
surface of a torus. This surface, plotted in Fig.~\ref{fig1},
can be parameterized as $T\left(u,v\right)=\left(\mathcal{R}\left(v\right)\cos u,\mathcal{R}\left(v\right)\sin u,r\sin v\right),$
where $\mathcal{R}\left(v\right)=R+r\cos v.$ Here the radius from
the center of the hole to the center of the torus tube is $r$, and
the radius of the tube is $R$. 

Introducing an imaginary Dirac mass, we will simulate an on-site gain
and/or loss term that induces non-Hermitian phases, similarly to the
flat case \citep{Wang,Nori}. The Dirac equation in $2+1$ dimensions
for a fermion of \emph{complex} mass $\mu=M+i\Gamma$ in a curved
background is described by a two-component spinor $\Psi(\mathbf{x})$
reads \citep{Collas_2019,Wald:1984rg} 
\begin{equation}\label{eq1}
\left(i\bar{\gamma}^{\nu}\nabla_{\nu}+\mu\right)\Psi(\mathbf{x})=0.
\end{equation}
 The space-time coordinates are labeled by $\mathbf{x}=\left(t,u,v\right)$,
the covariant derivative is given by $\nabla_{\nu}=\partial_{\nu}+\frac{1}{8}\omega_{\nu AB}\left[\gamma^{A},\gamma^{B}\right]$.
The spin connection $\omega_{\mu AB}=\eta_{AC}e_{\nu}^{\ \ C}\left(e_{\ \ B}^{\sigma}\Gamma_{\ \ \sigma\mu}^{\nu}+\partial_{\mu}e_{\ \ B}^{\nu}\right)$
is written in terms of the metric connection $\Gamma_{\sigma\mu}^{\nu}=\frac{1}{2}g^{\beta\nu}\left(\partial_{\mu}g_{\sigma\beta}-\partial_{\beta}g_{\sigma\mu}+\partial_{\sigma}g_{\mu\beta}\right).$
We are using $A,B,\ldots$ for local frame indices and $\mu,\nu,\ldots$
for coordinate indices. The metric of this space is $g_{\mu\nu}=\text{diag}\left(-1,\mathcal{R}\left(v\right)^{2},r^{2}\right)$,
with $u,\ v\in[0,2\pi)$. The \emph{vielbein} is given by the definition
$g_{\mu\nu}=\eta_{AB}e_{\mu}^{\ \ A}e_{\nu}^{\ \ B},$ where $\eta_{AB}=\text{diag}\left(-1,1,1\right)$.
We also defined the inverse \emph{vielbein} to satisfy $e_{\mu}^{\ \ A}e_{\ \ B}^{\mu}=\delta_{B}^{A}.$
We write the gamma matrices as $\bar{\gamma}^{\mu}=e_{\ \ A}^{\mu}\gamma^{A}$,
where we choose $\gamma^{0}=\sigma^{3},$ $\gamma^{1}=-i\sigma^{2},$
$\gamma^{2}=-i\sigma^{1},$ consistently with the anti-commutation
relations $\left\{ \bar{\gamma}^{\mu},\bar{\gamma}^{\nu}\right\} =-2g^{\mu\nu}$
\citep{Ye_and_Job2022}. By separating the time component from the
spatial ones in Eq. (\ref{eq1}), considering $M=0$, we obtain
$i\partial_{t}\Psi(t,u,v)=\mathcal{H}\Psi(t,u,v),$ where

\begin{equation}
\mathcal{H}=\left(\begin{array}{cc}
-i\Gamma & \mathfrak{D}^{-}\\
\mathfrak{D}^{+} & i\Gamma
\end{array}\right),\label{eq2}
\end{equation}
with 
\begin{equation}\label{eq3}
\mathfrak{D}^{\pm}=\pm\frac{1}{r}\left(\partial_{v}+\frac{1}{2}\frac{\mathcal{R}^{\prime}}{\mathcal{R}}\right)+\frac{i}{\mathcal{R}}\partial_{u}.
\end{equation}
The Hamiltonian of the system described by Eq.~$(\ref{eq1})$ obeys the symmetry relation $\left[\mathcal{H},\mathcal{PT}\right]=0$. For the parity and temporal reversal operators defined respectively, $\mathcal{P}=\sigma_x$ and $\mathcal{T}=-i\sigma_y K$, where $\sigma_x$ and $\sigma_y$ are the matrices of Pauli, and $K$ is the complex conjugate operator. This commutation relation guarantees that even with the non-Hermitian Hamiltonian we obtain real eigenvalues \cite{Bender,Bender2007}. 

 The solutions must be must be of the form \citep{Ye_and_Job2022},
when we isolate the $u$-dependence by expanding the spinors into
the Fourier series, 
\begin{equation}
\Psi\left(\mathbf{x}\right)=e^{-iEt}\sum_{m}\frac{e^{-imu}}{\sqrt{2\pi}}\left(\begin{array}{c}
\psi_{1}\left(v\right)\\
\psi_{2}\left(v\right)
\end{array}\right),\label{eq4}
\end{equation}
 where $m=\pm\frac{1}{2},\pm\frac{3}{2},...$ since we work with spin$\nicefrac{1}{2}$
field. This implies in a Schrodinger like eigenvalue problem $\mathcal{H}\psi\left(v,u\right)=E\psi\left(v,u\right),$where
we obtain 
\begin{align}\label{eq5}
-\left[\partial_{v}-\frac{r}{\mathcal{R}}\left(m+\frac{1}{2}\sin v\right)\right]\psi_{2}\left(v\right) & =r\left(E+i\Gamma\right)\psi_{1}\left(v\right),\\ \label{eq6}
\left[\partial_{v}+\frac{r}{\mathcal{R}}\left(m-\frac{1}{2}\sin v\right)\right]\psi_{1}\left(v\right) & =r\left(E-i\Gamma\right)\psi_{2}\left(v\right).
\end{align}

Squaring $\mathcal{H}$ and making the change of variables $x=\cos v$,
$x\in\left[-1,1\right],$ we obtain 
\begin{equation}\label{eq7}
\frac{1}{r^{2}}\left\{ -\partial_{v}^{2}-\frac{\mathcal{R}^{\prime}}{\mathcal{R}}\partial_{v}+V\left(v\right)-r^{2}\Gamma^{2}\right\} \left(\begin{array}{c}
\psi_{1}\\
\psi_{2}
\end{array}\right)=E^{2}\left(\begin{array}{c}
\psi_{1}\\
\psi_{2}
\end{array}\right),
\end{equation}
 where 
\begin{align}\label{eq8}
V\left(v\right) & =\frac{1}{\mathcal{R}^{2}}\left(m^{2}r^{2}+\sigma^{3}rm\mathcal{R}^{\prime}+\frac{1}{4}\mathcal{R}^{\prime2}-\frac{1}{2}\mathcal{R}\mathcal{R}^{\prime\prime}\right)\\ \label{eq9}
\mathcal{R}\left(v\right) & =R+r\cos v.
\end{align}

It is important to discuss at this point the Gaussian curvature for the torus. The Gaussian curvature can be straightforwardly calculated from the metric given previously, so that we obtain
\begin{equation}
    K=\frac{\cos(v)}{r[R+r\cos(v)]}.
\end{equation}

\begin{figure}
    \centering
    \includegraphics[scale=0.7]{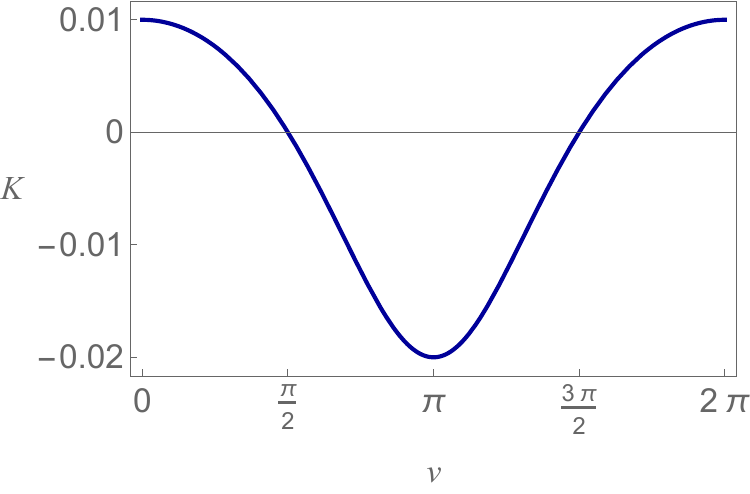}
    \caption{Torus Gaussian curvature}
    \label{gaussiancurvaturefig}
\end{figure}

From the fig. (\ref{gaussiancurvaturefig}) we can clearly see that there are four points of interest that deserve a few comments. These points of interest are related to the sign of the Gaussian curvature. At $v=0$ the Gaussian curvature assumes a maximum positive value, while at $v=\pi$ it reaches its mininum negative value. And at $v=\pi/2$ and $v=3\pi/2$ the Gaussian curvature vanishes. 

Now we are in position to investigate the non-hermitian phase transitions within the system by solving the eq. (\ref{eq7}) in order to study the behaviour of the exceptional points. Analytical solutions could not be found for the present case, thus we perform the non-Hermitian phase transition investigation numerically, by applying the finite elements methods in order to obtain the spectrum of eigenvalues of the problem.
\section{Results}

The non-Hermitian phase transition induced by imaginary Dirac mass was studied in the geometry of the sphere with the emergence of an exceptional point \cite{Ygor2021}. On the torus surface, we find the emergence of multiple non-Hermitian phase transitions. The distinction between the non-hermitian phase transition in the sphere and in the torus are, in a certain sense, expected, since the two geometries are topologically distinct. 

Numerically, we obtain the spectrum of eigenvalues of the problem. We observe the presence of exceptional points that clearly delimits the region where $\mathcal{PT}$ symmetry breaks. The set of real eigenvalues passes through the exceptional point and becomes complex. We observed that at the limit where $r$ is much smaller than $R$, increasing $\Gamma$ we observe that the first exceptional point approaches to zero. The growth of $r$ induces the emergence of more exceptional points in the problem. This can be linked to the behavior of the Ricci scalar in relation to this parameter.

We can investigate the non-Hermitian phase transitions here in two ways. The first way is by setting a particular configuration of the torus and varying the magnitude of the imaginary component of the mass parameter. The second way is by setting a particular value for the imaginary component of the mass parameter and change the torus configuration itself. In the first way we are interested in studying the sole influence of the imaginary mass in the non-Hermitian phase transitions while in the second way we are keen to understand the role played by the geometry in the non-Hermitian phase transitions. 

\begin{figure}[t]
\includegraphics[width=0.9\columnwidth]{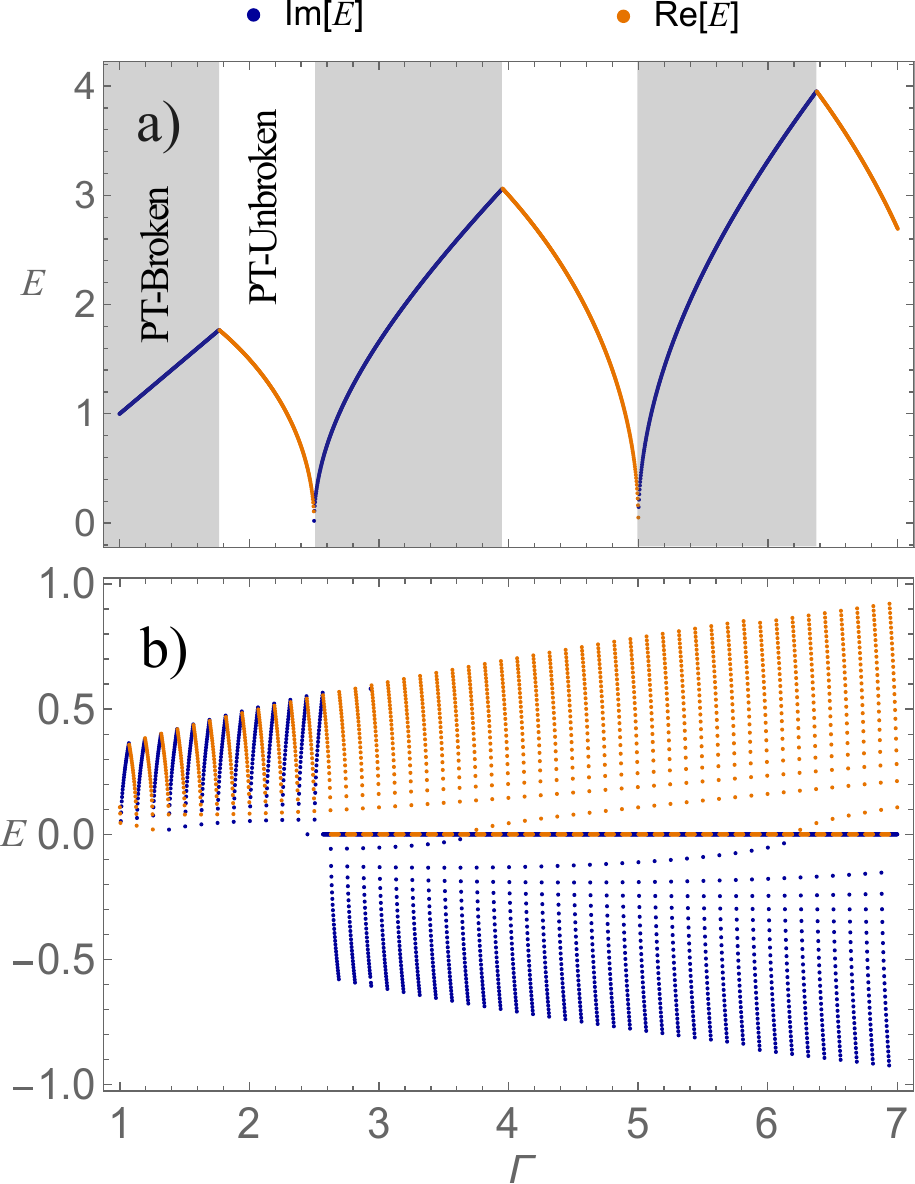}
\caption{Multiple non-Hermitian phase transitions of $E \times \Gamma$ for the first eigenvalue of Eq.~(\ref{eq7}) with $R=10$ and $m=1/2$. Shows the phase transitions in the cases of (a) $r=0.4$ and (b) $r=8$. The blue dots describe the imaginary part of $E$ which describe the $\mathcal{PT}$- broken regions, and the orange dots are the real parts of $E$ representing the $\mathcal{PT}$-Unbroken regions. \label{fig2}}
\end{figure}

Let us address the initially the first way. In Fig.~(\ref{fig2}), we show the behavior of the first eigenvalue of Eq.~(\ref{eq7}) with $R=10$ and the $m=1/2$.  Due the symmetry of the problem, we can restrict our analysis by setting values of $v = \left[0,2\pi \right]$. In Fig.~(\ref{fig2}a) we have set $r=0.4$, thus describing a ring-like configuration, and we can notice that the ground state eigenvalue starts entirely imaginary at $\Gamma=1$ and it grows almost linearly up to $\Gamma=1.8$. After $\Gamma=1.8$ the ground state eigenvalue becomes entirely real but now exhibiting a decreasing behavior. The ground state eigenvalue continues to decrease until it becomes zero and it starts increasing again but now entirely complex. This pattern repeats as we increase the value of $\Gamma$ and the absolute values of the energy of the system increases as well as the value of $\Gamma$ becomes larger, as expected. It becomes clear, therefore, that the magnitude of the imaginary component of the mass controls weather the system exhibit a PT-broken or a PT-unbroken phase. 

\begin{figure}[t]
	\includegraphics[width=1\columnwidth]{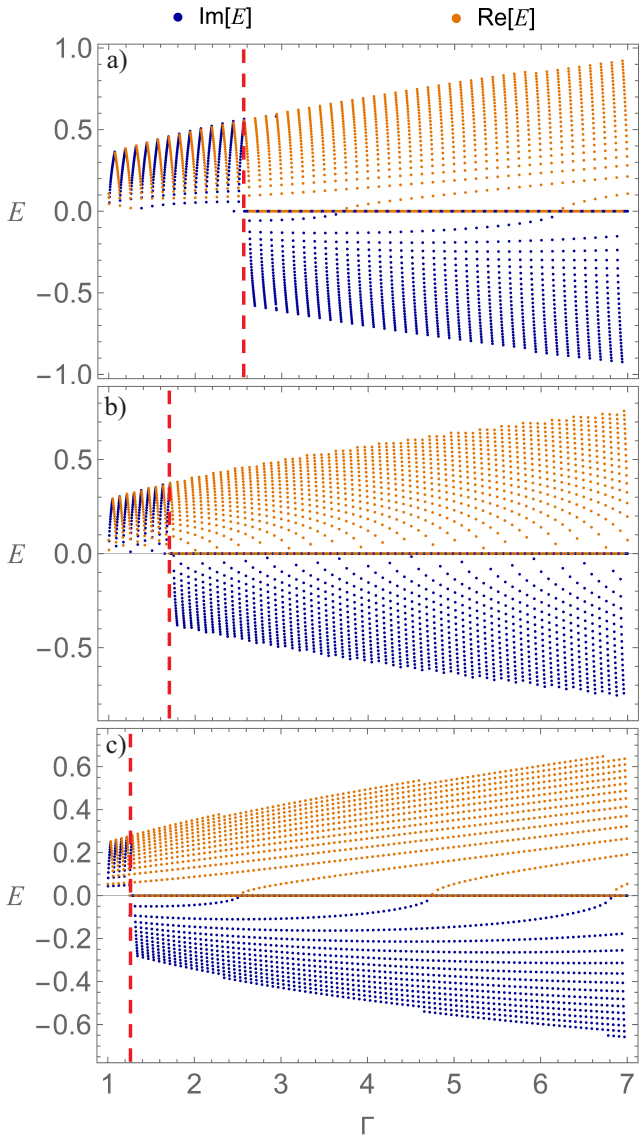}
	\caption{Multiple phase transitions for optimal proportion regions with ratio $R/r=1.25$ and $m=1/2$. a) $R=10$ and $r=8$. (b) $R=15$ and $r=12$. (c) $R=20$ and $r=16$. The red line shows the change in behavior of the eigenvalues.   \label{fig3} }
\end{figure}

\begin{figure}[t]
	\centering
	\includegraphics[width=1\columnwidth]{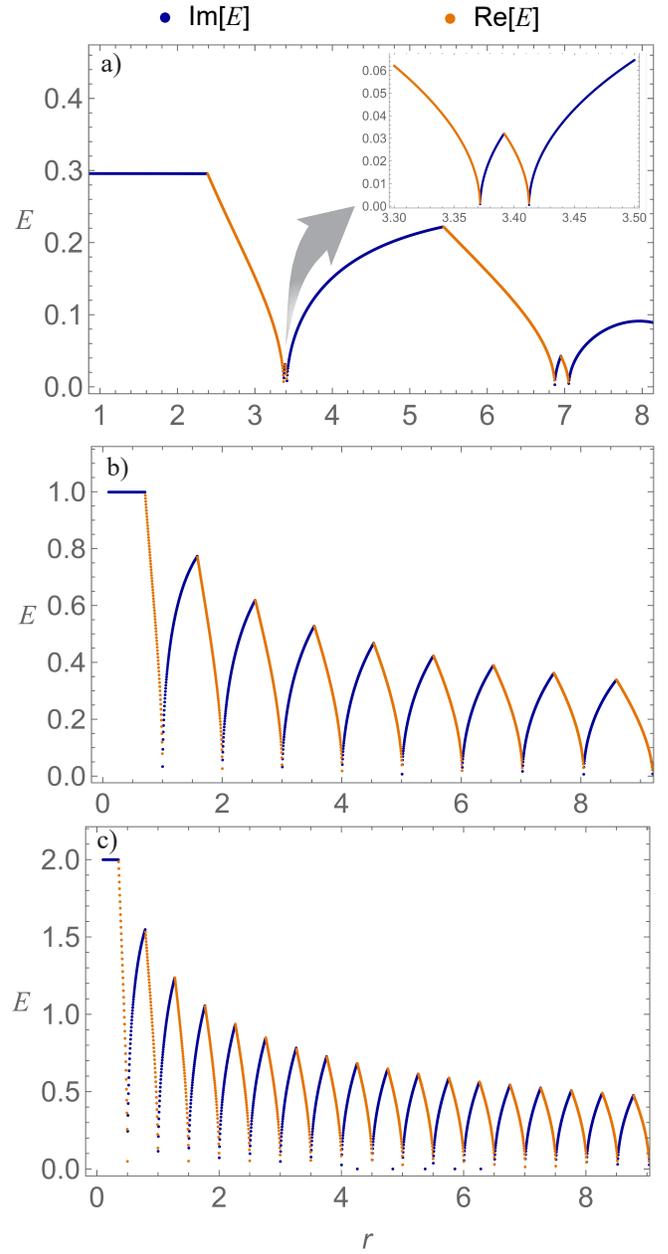}
	\caption{Multiple phase transitions for $E \times r$ with $R=10$ and $m=1/2$. (a) $\Gamma=0.3$. (b) $\Gamma = 1.$ (c) $\Gamma = 2$.\label{fig4} }
\end{figure}

In Fig.~(\ref{fig2}b) we have set $r=8$, being therefore in a much more torus-like configuration, in comparison with the ring-like configuration considered in Fig.~(\ref{fig2}a). In such torus-like configuration it is clear that, similarly to the ring-like configuration, the magnitude of $\Gamma$ controls the phase of the system and we can also notice that, as we increase the value of $\Gamma$, the number of phase transitions in the torus-like configuration is much grater than in the ring-like one. However, it is important to highlight here that a whole new kind of behavior emerges in such torus-like configuration. As we can see in Fig.~(\ref{fig2}b), up to a certain value of $\Gamma$, around $\Gamma=2.6$, the energy eigenvalues are entirely positive, regardless being real or imaginary. However, for values of $\Gamma$ greater than $\Gamma=2.6$, all the real eigenvalues of the energy which are real are positive while all the energy eigenvalues which are imaginary are negative. This means that the system presents a gain or loss, there is a change from gain/loss to loss/gain or vice versa \cite{Yuto_Ashida2020}.

An interesting fact that must be pointed out is that such drastic change in the sign of the imaginary energy eigenvalues in the torus-like configuration only happens for a certain torus configuration. In Fig.~(\ref{fig2}b) we have considered $R=10$ and $r=8$, so that the ratio $R/r=1.25$ sets the configuration in which this sign change in the imaginary energy eigenvalues emerge.

In Fig.~$(\ref{fig3})$ we investigate the sign change in the imaginary energy eigenvalue for three torus-like configuration, namely, $R=10$ and $r=8$, $R=15$ and $r=12$, and finally $R=20$ and $r=16$. In all three configurations the ratio $R/r=1.25$. For the configuration depicted in Fig.~$(\ref{fig3}a)$, i.e. $R=10$ and $r=8$, we can see that the sign change occurs when $\Gamma=2.6$, as pointed out previously. For the case Fig.~$(\ref{fig3}b)$, when $R=15$ and $r=12$, a small value of $\Gamma$ sets the sign change, namely, $\Gamma=1.7$. And for Fig.~$(\ref{fig3}c)$, case where $R=20$ and $r=16$, an even small value of $\Gamma=1.22$ sets the sign change. The comparison between the three cases allows us to conclude that the sign change point shifting to the left could be related to the increase in the value of $r$. Since the ratio $R/r$ is kept the same for all cases, but the $r$ increases from Fig.~$(\ref{fig3}a)$ to Fig.~$(\ref{fig3}c)$, then the magnitude of $\Gamma$ where the sign change in the imaginary energy eigenvalues could be inversely related to the value of $r$, and therefore to the value of the Gaussian curvature, which depends only on $r$, for a fixed angle. Optimal torus proportions, expressed in terms of the ratio $R/r$, were already reported in the literature (see f.e. \cite{Liu_Colossal2002}).  

In Fig.~$(\ref{fig4})$, we show the multiple phase transitions that emerge with the change of $r$, for the simulations we use $R=10$ and $m=1/2$. In Fig.~(\ref{fig4}a), we use $\Gamma = 0.3$, in which small regions of phase transitions were observed. Initially, the eigenvalues are complex, remaining fixed up to $r=2.4$, and the eigenvalues become real going to zero at $r=3.37$ (see insert Fig.~\ref{fig4}a). In the insert we also find a small complex/real phase transition, and then there are more phase transitions, these being long and short. In Fig.~(\ref{fig4}b) and Fig.~(\ref{fig4}), we observe the increase in phase transitions with the increase in $\Gamma$, where we use $\Gamma=1$ and $\Gamma=2$, respectively.

\section{Conclusions}

In this paper we have investigated the arising of non-hermitian phase transitions on a quantum torus surfaces. We considered a single fermion whose dynamics is governed by the Dirac equation confined to move on a quantum torus surface. The effects of the geometry are take into account by using the tetrad formalism and the spin connection. The Dirac equation gave rise to two coupled first-order differential equations for each spinor component. The eigenvalues and eigenfunctions for each spinor component were computed numerically and the non-hermitian phase transitions were investigated in terms of the geometric features of the torus and the magnitude of the imaginary component of the mass.

We have shown numerically that the surface of the torus with the non-Hermiticity induced from the Dirac imaginary mass term presents multiple non-Hermitian phase transitions. We find the phase transitions $\mathcal{PT}$ observed by changing the imaginary mass parameter $\Gamma$, as well as through the radii $R$ and $r$ of the torus. In the ring limit, when $r \ll R$, we determine that few non-Hermitian phase transitions occur. However, we observe that the transitions increase considerably when $r$ approaches $R$.
In addition, we highlight the non-Hermitian phase transition regions for the conditions of the torus radii of optimal proportion, in which the increase in radii in the same proportion of $R/r = 1.25$ induces a change in the behavior of the eigenvalues of $E$, characterizing the on-site gain or loss of the torus surface.
This investigation of non-Hermitian phases in curved space allows us to analyze non-Hermitian physics and its properties in new geometries by searching for new effects.

\section*{Acknowledgements}
JF would like to thank the Funda\c{c}\~{a}o Cearense de Apoio ao Desenvolvimento Cient\'{i}fico e Tecnol\'{o}gico (FUNCAP) under the grant PRONEM PNE0112-00085.01.00/16 for financial support, the CNPq under the grant 304485/2023-3, Gazi University for the kind hospitality, Alexandra Elbakyan and Sci-Hub, for removing all barriers in the way of science .

%\begin{appendix}

%\section{Spectrum}\label{app.A}
%Before we use the Eq.(\cite{eq4}) its possible to define the spinor as $$\Psi\left(\mathbf{x}\right)=e^{-iEt}\left(\begin{array}{c}
%f_{1}\left(u\right)g_{1}\left(v\right)\\
%f_{2}\left(u\right)g_{2}\left(v\right)
%\end{array}\right).$$ To the down component \begin{align}
%\frac{1}{r^{2}}\frac{1}{g_{2}\left(v\right)}\left[-\partial_{v}^{2}g_{2}\left(v\right)-\frac{\mathcal{R}^{\prime}}{\mathcal{R}}\partial_{v}g_{2}\left(v\right)\right]f_{2}\left(u\right) & +\nonumber \\
%+\frac{1}{r^{2}}\left[\frac{1}{\mathcal{R}^{2}}\left(-r^{2}\partial_{u}^{2}-ir\mathcal{R}^{\prime}\partial_{u}+\frac{1}{4}\mathcal{R}^{\prime2}-\frac{1}{2}\mathcal{R}\mathcal{R}^{\prime}\right)-r^{2}\Gamma^{2}\right]f_{2}\left(u\right) & =E^{2}f_{2}\left(u\right)
%\end{align}

%\end{appendix}

\bibliography{Reference}

\end{document}